\documentclass[12pt]{article}

\usepackage{scicite}

\usepackage{times}

\usepackage{amsmath}
\usepackage{newtxmath}
\usepackage{amsfonts}
\usepackage{amssymb}
\usepackage{graphicx}

\usepackage[version=3]{mhchem} 
\usepackage[detect-weight=true, detect-family=true]{siunitx} 

\usepackage[latin1]{inputenc}
\usepackage{xspace}
\usepackage{color}

\usepackage[hidelinks]{hyperref}
\usepackage{bm}
\usepackage{mathtools}

\DeclarePairedDelimiter{\abs}{\lvert}{\rvert}
\DeclarePairedDelimiter\ket{\lvert}{\rangle}

\newenvironment{sciabstract}{%
\begin{quote} \bf}
{\end{quote}}

\title{Tip-enhanced strong coupling spectroscopy, imaging, and control of a single quantum emitter}

\author{Kyoung-Duck Park,${}^{1\dagger,\ast}$ Molly A. May,${}^{1\dagger}$ Haixu Leng,${}^{2}$  Jiarong Wang,${}^{1}$ \\ Jaron A. Kropp,${}^{2}$ Theodosia Gougousi,${}^{2}$  Matthew Pelton,${}^{2\ast}$ and  Markus B. Raschke${}^{1\ast}$\\
\\
\normalsize{${}^{1}$Department of Physics, Department of Chemistry, and JILA,} \\\normalsize{University of Colorado, Boulder, CO 80309, USA}\\
\normalsize{${}^{2}$Department of Physics, University of Maryland,} \\ 
\normalsize{Baltimore County, Baltimore, MD 21250, USA}\\
\\
\normalsize{$^\dagger$These authors contributed equally to this work.} \\
\normalsize{$^\ast$To whom correspondence should be addressed;} \\ 
\normalsize{E-mail: markus.raschke@colorado.edu (M.B.R.), mpelton@umbc.edu (M.P.),} \\ 
\normalsize{kyoungduck.park@colorado.edu (K.-D.P.)}
}

\date{}

\begin{document} 


\baselineskip24pt


\maketitle



\begin{sciabstract}
Optical cavities can enhance and control light-matter interactions. This has recently been extended to the nanoscale, and with single emitter strong coupling regime even at room temperature using plasmonic nano-cavities with deep sub-diffraction-limited mode volumes. However, with emitters in static nano-cavities, this limits the ability to tune coupling strength or to couple different emitters to the same cavity. Here, we present tip-enhanced strong coupling (TESC) spectroscopy, imaging, and control.  Based on a nano-cavity formed between a scanning plasmonic antenna-tip and the substrate, by reversibly and dynamically addressing single quantum dots (QDs) we observe mode splitting $>$ 160 meV and anticrossing over a detuning range of $\sim$100 meV, and with sub-nm precision control over the mode volume in the $\sim$10$^{3}$ nm$^{3}$ regime. Our approach, as a new paradigm of nano-cavity quantum-electrodynamics near-field microscopy to induce, probe, and control single-emitter plasmon hybrid quantum states, opens new pathways from opto-electronics to quantum information science.

\end{sciabstract}



\noindent Single quantum emitters in solids in the form of quantum dots (QDs), nitrogen vacancy centers, or engineered defects as artificial atoms have emerged as promising platforms for quantum sensing, metrology, and information processing \cite{claudon2010, kurtsiefer2000, he2015, palacios2016}. These quantum emitters can now be controlled using optical cavities to enhance the coupling strength $g$ between emitters and cavity photons to the point where it exceeds the rates of quantum decoherence in the system. However, the diffraction limit restricts the mode volume $V$ and thus the coupling strength $g\propto1/\sqrt{V}$, which has so far required operation at cryogenic temperatures to overcome decoherence \cite{yoshie2004, reithmaier2004, khitrova2006, hennessy2007, vasa2013}.

As an alternative approach, plasmonic cavities with nanoscale mode volumes provide a promising platform for ultra-compact cavity quantum electrodynamics (cQED) systems even at room temperature \cite{pelton2015, bharadwaj2006, seelig2007, fofang2011, chen2013, zengin2015, chikkaraddy2016, melnikau2016, santhosh2016, liu2017, chen2011}.
Specifically, QDs or molecules coupled strongly to plasmonic nano-cavities give rise to a plexcitonic state as a hybrid state of a plasmon and an exciton \cite{manjavacas2011}.

The observation of these plexcitonic states from single emitters was recently demonstrated with Rabi splitting in scattering spectra \cite{chikkaraddy2016, santhosh2016} even at room temperature. However, in contrast to photoluminescence (PL), the discrimination of strong coupling from competing Fano-like interference effects can still be ambiguous in scattering \cite{leng2018}. In addition, once fabricated, the static QD nanocavity devices constrain tunability and control. Further, due to their nanoscopic dimensions, the details of field confinement in these structures are not accessible to conventional microscopy. This limits the ability to measure, optimize, and control coupling and dissipation. Therefore, in order to expand the utility of this nano-cavity approach, a nano-imaging based implementation of plasmonic strong coupling is desired. 

Here, we demonstrate tip-enhanced strong coupling (TESC) spectroscopy, imaging, and control based on scanning probe microscopy.
A resonant plasmonic cavity with nanoscale mode volume of $V/\lambda^3$ $\leq$ 10$^{-6}$ is deterministically formed between a nano-optical antenna-tip  and a metal mirror substrate. With this qualitatively new approach, we scan, locally address, and probe single QDs at room temperature achieving clear, high-contrast plexcitonic PL with Rabi splitting up to 163 meV. We probe different QDs with exciton energies detuned from the plasmonic cavity over a range of $\sim$100 meV, thereby demonstrating anticrossing of the upper and lower polariton modes. In addition, based on the precise sub-nm distance control of the tip, we spatially map and reversibly modulate the coupling strength over a range of up to 140 meV.

\begin{figure}
\center
\includegraphics[width=16cm]{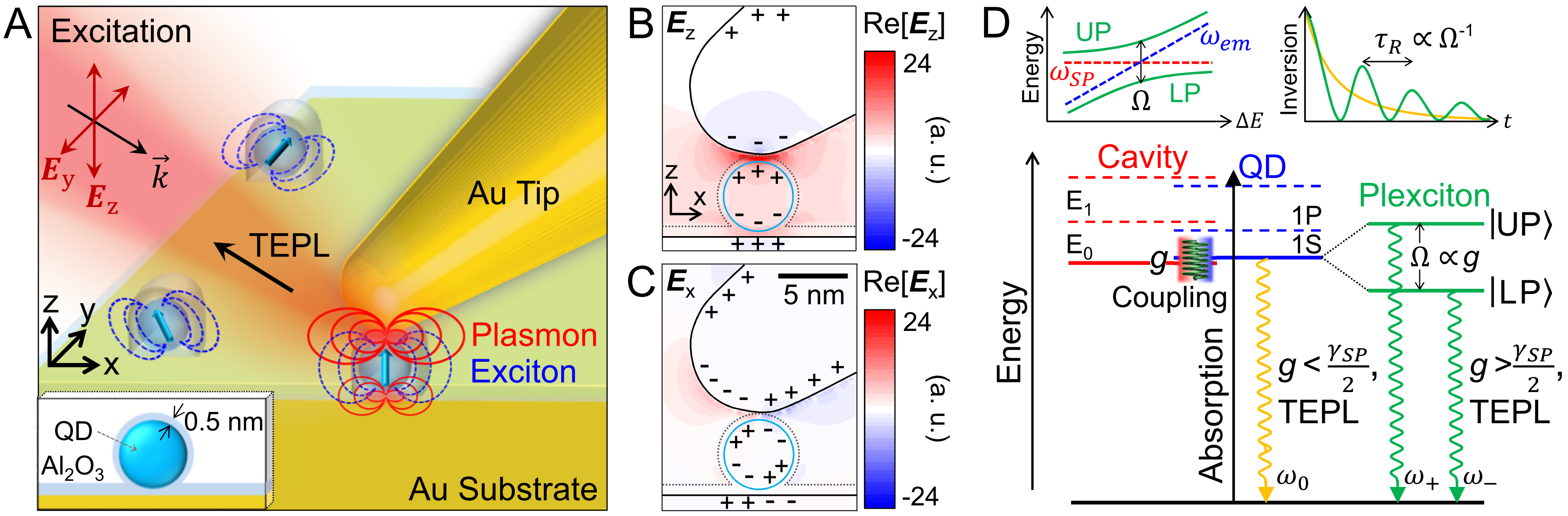}
\caption{
\textbf{Tip-enhanced strong coupling (TESC) spectroscopy and energy diagram for the plasmon and exciton in the weak and strong coupling regime.} 
(A) The strongly confined $|\textbf{E}_z|$ fields in a 0.5 nm dielectric capping layer (Al$_2$O$_3$) of a single isolated quantum dot (CdSe/ZnS) with a tilted Au tip induce coupling between the plasmon and exciton. Simulated \textit{out-of-plane} (B) and \textit{in-plane} (C) optical field distributions in the plasmonic-cavity shown in (A). 
(D) Energy diagram for the plasmonic-cavity (red), quantum dot (blue), and upper and lower polariton states (green) with photoluminescence energy in the weak (orange) and strong (green) coupling regimes. 
When the coupling exceeds system losses, the split polariton states emerge and the system begins to undergo Rabi splittings and Rabi oscillations as illustrated above.
}
\label{fig1}
\end{figure}

This work extends the related work by Gro{\ss} et al. \cite{grob2018}, which employed an engineered slot antenna on a tip to achieve strong coupling of single QDs near the slot. However, the slot antenna limits the cavity mode volume and spatial resolution that can be obtained, and does not enable dynamical control of the mode volume. Instead, in TESC we employ the \textit{inverse} structure of a conical optical antenna nano-tip to form a  nano-cavity between the tip and a metal mirror substrate. This eliminates the need to generate a nano-gap on the tip itself (Fig.~\ref{fig1}) and provides for both a greatly simplified and generalizable approach for active control of the confined nano-cavity field over distances down to the atomic scale. The resulting improved signal quality in TESC with unambiguous two-peak splitting, anti-crossing, and dynamical sub-nm nano-cavity control opens a new field of nano-cavity cQED.


TESC corresponds to a new paradigm of near-field optical microscopy. While all previous near-field optical microscopy efforts have focused on a minimally invasive tip-sample interaction to obtain an unperturbed spectroscopic signature \cite{atkin2012}, TESC  moves from a traditionally nonperturbative to a strongly perturbative regime in which the tip acts as a nano-cavity to induce, probe, and control new quantum hybrid states.

TESC is the optical analog to recent advances in scanning tunneling microscopy (STM) based on strongly perturbing tip-induced electric or spin interactions to induce and probe novel quantum magnetic or quantum coherence effects in a localized and tunable fashion \cite{yan2015, willke2018, baumann2015}. 

As the quantum emitter, we use single isolated CdSe/ZnS QDs, which are dropcast onto a flat, template-stripped Au surface that is coated with a thin Al$_2$O$_3$ layer. 
The QDs are protected against photo-oxidation by another ultrathin 0.5 nm Al$_2$O$_3$ capping layer and are characterized by atomic force microscopy (AFM) to confirm even dispersion of the particles (Supplementary Materials I, II and Fig. S1).
The plasmonic Au tips are etched electrochemically \cite{neacsu2005}.
Tip-enhanced photoluminescence (TEPL) spectra are recorded under continuous wave (632.8 nm, $\leq$ 1 mW) resonant tip-surface plasmon polariton (SPP) and QD-exciton excitation, as illustrated in Fig. 1A  (Supplementary Materials III). 
The tilted tip (35$^\circ$ with respect to the sample surface in this case) controls the tip-SPP resonance frequency and maximizes the field enhancement, as shown recently \cite{park2018polarization}.
In the resulting plasmonic cavity, the corresponding SPP mode is most strongly bound in $|\textbf{E}_z|$, as confirmed by finite-difference time domain (FDTD) simulations (Fig. 1B-C).
Sample scanning and tip-sample distance are controlled with $\sim$0.2 nm precision using a shear-force AFM.
All experiments are performed at room temperature.

Fig. 1D shows a schematic energy diagram for the plasmon (red), exciton (blue), and their hybridized plexciton (green) with upper $\ket{\mathrm{UP}}$ and lower polariton $\ket{\mathrm{LP}}$ states.
When the coupling strength $g$ between the cavity plasmon and the QD exciton exceeds the SPP loss rate, $\gamma_{SP}$, i.e., $g>\frac{\gamma_{SP}}{2}$, quantum mixed states of the plexciton give rise to Rabi splitting in the TEPL spectra, or equivalently, a reversible Rabi oscillation between the plasmon and exciton.
In contrast, in the weak coupling regime ($g<\frac{\gamma_{SP}}{2}$), the usual primarily radiative relaxation with enhanced PL intensity (orange) attributed to the plasmonic Purcell effect are observed at the exciton frequency \cite{park2016tmd}.

\begin{figure}
\center
\includegraphics[width=16cm]{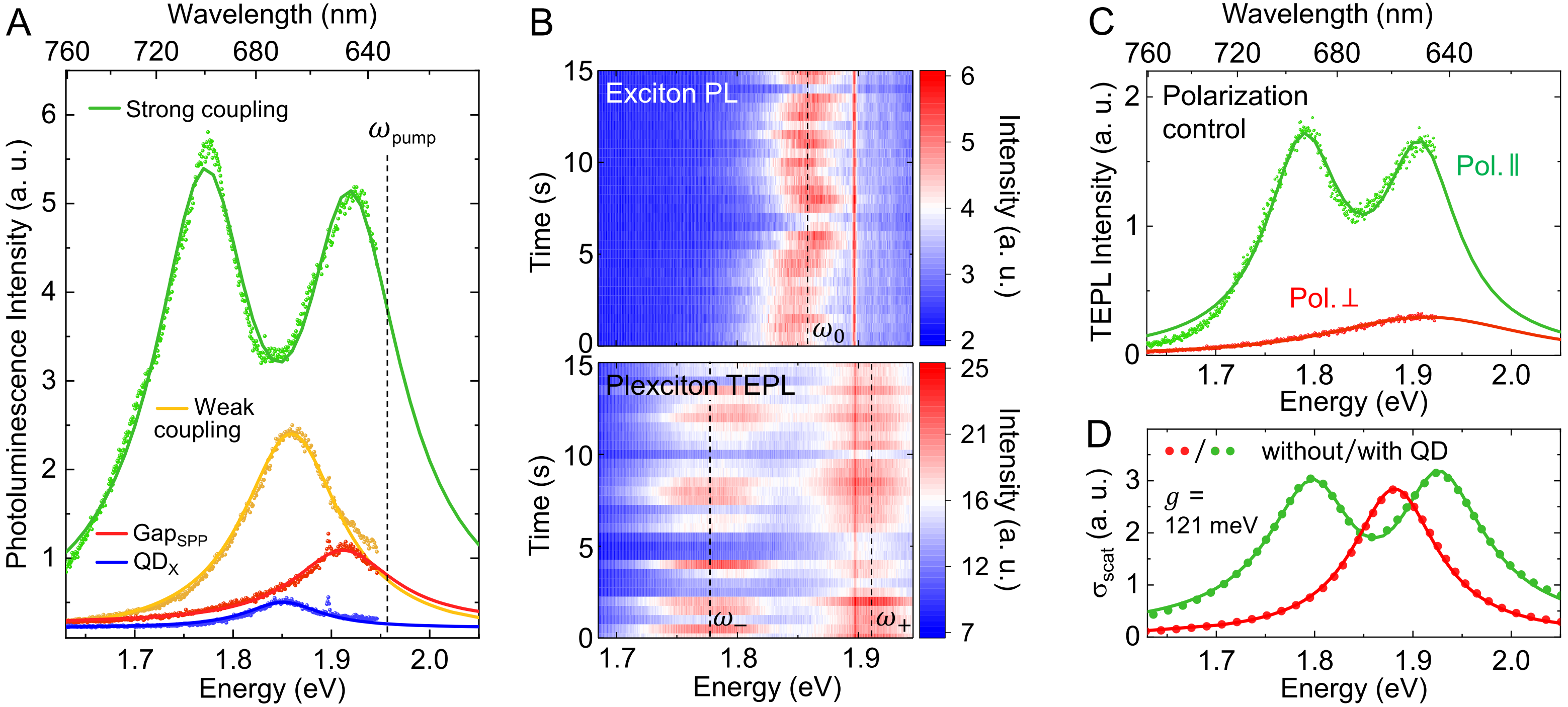}
\caption{
\textbf{Tip-enhanced plexciton photoluminescence at room temperature in the strong coupling regime.} 
(A) PL spectra of the gap plasmon (red), QD exciton (blue), the weakly coupled plasmon-exciton mode (orange), and the strongly coupled plexciton mode (green).  
(B) PL evolution of the uncoupled (top) and the strongly coupled (bottom) single QD. 
(C) TEPL spectra for a polarization parallel (green) and perpendicular (red) with respect to the tip.
(D) Finite element method (FEM) simulation of scattering spectra for the plasmonic-cavity without QD (red) and with a single QD (green).
In (A), (C), and (D), the dots and lines indicate the measurement (or simulation) data and the corresponding model fits, respectively.
}
\label{fig2}
\end{figure}

We first characterize the emission properties of the uncoupled QD excitons in the far-field, as shown by the PL spectrum (blue) in Fig. 2A.
Independently, we characterize the tip-surface nano-cavity through its tip-SPP emission \cite{kravtsov2014} (red), with the tip retracted to $\sim$3 nm from a clean Au mirror substrate without QDs.
From fitting to a Lorentzian lineshape function, corresponding peak energies ($\omega_{QD}$ = 1.856 eV and $\omega_{SP}$ = 1.912 eV) and linewidths ($\gamma_{QD}$ = 0.097 $\pm$ 0.002 eV and $\gamma_{SP}$ = 0.126 $\pm$ 0.002 eV) are derived (Supplementary Materials IV and Fig. S2).
We note that $\omega_{SP}$ is expected to red-shift slightly by $<$2\% in the presence of the QD due to modification of the effective dielectric environment in the nano-cavity \cite{jensen1999}. This is beneficial for increasing spectral overlap of SPP and QD resonances.
We then scan and position the tip directly over an individual QD and measure its TEPL response coupled to the plasmonic cavity.
In a series of experiments for different QDs with the same tip, we observe TEPL spectra across a wide range of coupling strengths, due to the random dipole orientation of the different QDs on the mirror substrate as shown in Fig. 2A. 
Weak coupling leads only to enhanced PL with spectral broadening, with $\omega_{weak}$ = 1.860 eV and $\gamma_{weak}$ = 0.118 eV (orange), essentially identical to the unperturbed QD exciton peak.
In contrast, in the strong coupling regime (green), for QDs with an \textit{out-of-plane} transition dipole moment (TDM) with respect to the surface, a qualitatively different behavior with peak splitting and intensity enhancement is observed.

We fit the observed TEPL intensity spectra $I_{PL}(\omega)$ using a coupled harmonic oscillator model in the Weisskopf-Wigner approximation for an impulsively excited emitter, given by \cite{cui2006} 
\begin{equation}
\label{Model}
I_{PL}(\omega) = \frac{\gamma_{QD}}{2\pi}\abs*{\frac{\gamma_{SP}/2-i(\omega-\omega_{SP})}{\{(\gamma_{SP}+\gamma_{QD})/4-i(\omega_{QD}-\omega_{SP})/2-i(\omega-\omega_{QD})\}^2+\Omega^2}}^2.
\end{equation}

Here, $\gamma_{SP}$ and $\gamma_{QD}$ are the decay rates, $\omega_{SP}$ and $\omega_{QD}$ are the resonance frequencies of plasmon mode and QD, respectively, and $\Omega$ is the vacuum Rabi frequency (for details see Supplementary Materials V).
The corresponding coupling strength $g$ can then be obtained from the fit parameters (Supplementary Materials VI) as
\begin{equation}
\label{RabiFreq}
g = 2\sqrt{\Omega^2-\frac{(\omega_{QD}-\omega_{SP})^2}{4}+\frac{(\gamma_{SP}-\gamma_{QD})^2}{16}}.
\end{equation}

For the example shown in Fig. 2A, we obtain a coupling strength of $g \sim$ 143 meV $>$ $\frac{\gamma_{SP}}{2}$, i.e., well in the strong coupling regime.
More extended theoretical models which take into account both plasmonic and QD losses reproduce the same spectral behavior of our experimental data \cite{laussy2009, torma2014}.

It is worth noting that the Raman peak of Al$_2$O$_3$ ($\bar{\nu}$ = 1.897 eV) observed in the far-field PL spectrum is not discernible in the TEPL spectra. 
This is because of the sparse distribution of QDs, compared to the large area of Al$_2$O$_3$ coverage that contributes to the far-field signal.

Fig. 2B (top) shows the PL evolution of the uncoupled QD, which exhibits the spectral diffusion and blinking behavior that is generally observed for different QDs.
The plexciton TEPL spectra in Fig. 2B (bottom) reveal a similar intensity fluctuation and no significant changes in coupling strength, which confirms that the tip-sample distance is precisely controlled in our experiments. 
Additionally, the result shows that the typical blinking due to, e.g., non-radiative Auger recombination and/or thermalization processes \cite{efros2016}, still occurs in the strong coupling regime.

As an important control experiment, we can vary the field enhancement in the $|\textbf{E}_z|$ direction through rotation of the incident laser polarization. 
For a QD with an \textit{out-of-plane} TDM, for polarization parallel with respect to the tip axis, plexciton formation and emission (green) is induced, as shown in Fig. 2C. 
In contrast, for perpendicular polarization, only the tip plasmon PL is observed (red).
This clear contrast concurs with 3D FDTD simulations for the given experimental conditions, which show that the excitation polarization parallel with respect to the tip gives rise to an $\sim$800-fold field intensity enhancement compared to polarization perpendicular with respect to the tip (see Fig. S3), and which confirm the well defined nature of the plasmonic nano-cavity.

We next model the spectra and the observed Rabi splitting based on finite element method (FEM) calculations \cite{wu2010}.
In these models, the QD is treated as a dielectric nanoparticle with oscillator strength $f \sim$ 0.8, a value that yields the best agreement with experimental observations (Fig. 2A and Fig. 2C) and consistent with recent work \cite{santhosh2016}.
Approximating the plasmonic antenna-tip as a finite length ellipsoid, the resulting simulated scattering spectrum, as shown in Fig. 2D, exhibits the expected splitting of the hybridized state with $g \sim$ 121 meV, in good agreement with the experiment (Supplementary Materials VII).

\begin{figure}
\center
\includegraphics[width=12cm]{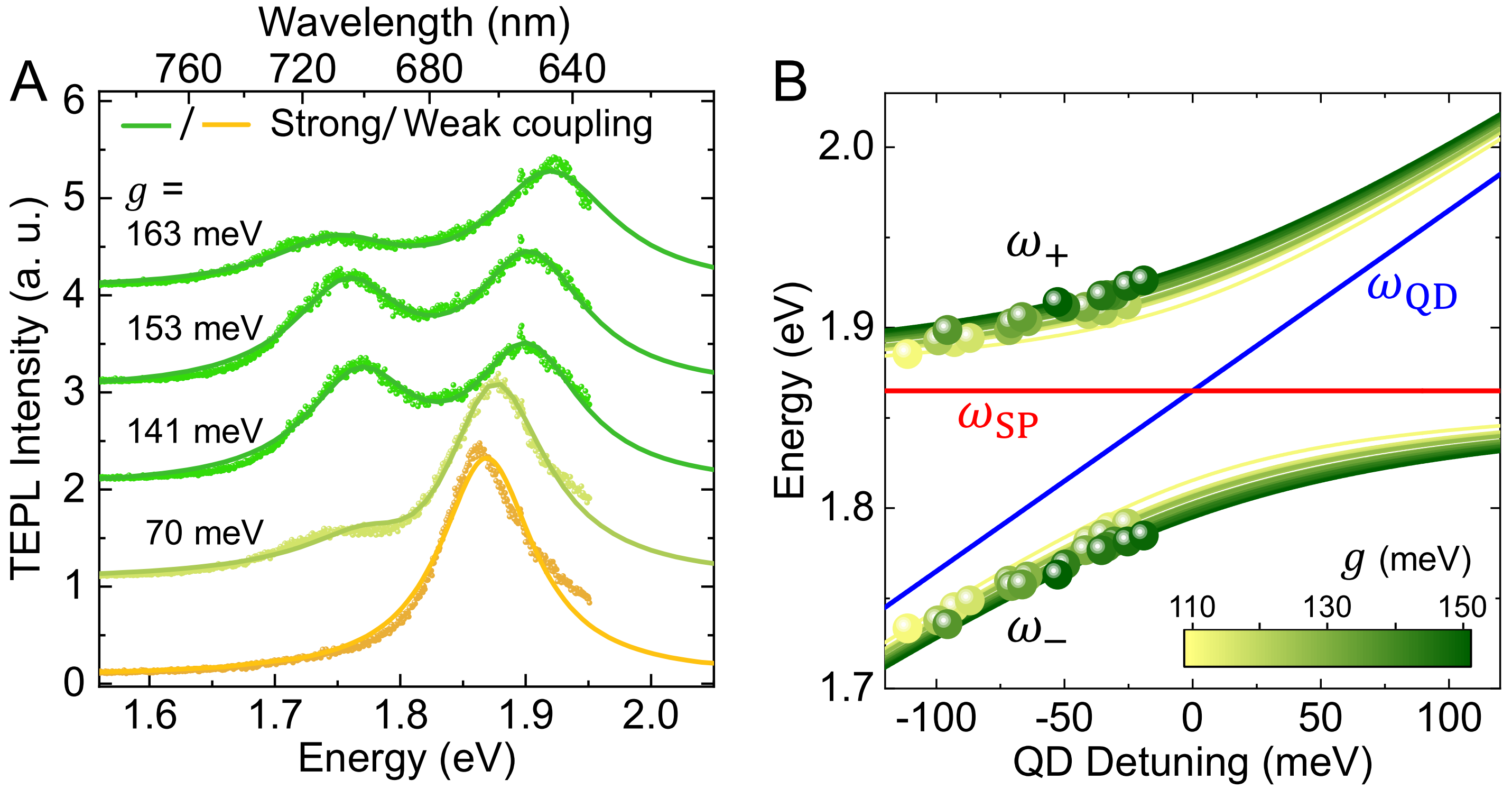}
\caption{
\textbf{TESC spectra with increasing coupling strength and plexciton energy diagram with QD detuning.} 
(A) TEPL spectra of different single QDs with variation in coupling strength $g$ and Rabi frequency.
(B) Polariton energies from model fits (circles) and anticrossing curves from model calculations (lines) for each of the measured TEPL spectra of 21 different QDs. 
The expected surface plasmon (red) and QD (blue) detuning dependence is obtained from averaged values from the modeled 21 spectra.
}
\label{anticrossing}
\end{figure}

As the tip interacts with different QDs in a large-area scan, we observe variations in plexciton coupling strength ranging from 70 meV (at the threshold for strong coupling) to 163 meV (well into the strong coupling regime), as shown in Fig. 3A. 
Since CdSe QDs have a TDM ($\mu_{QD}$) oriented perpendicular to the crystallographic \textit{c}-axis of the nanocrystal \cite{empedocles1999}, this variation can be described by the different orientations of QDs, which modify the coupling strength with the cavity $|\textbf{E}_z|$ field \cite{andersen2011}, given by $g$ $\approx$ $|\mu_{QD}||\textbf{E}_z|$cos$\thetaup$, where $\thetaup$ denotes the angle between the TDM and the surface-normal.
In addition to the QD orientation, the inevitable inhomogeneities in local QD environment, e.g., different Al$_2$O$_3$ layer thickness and roughness of the Au film, also give rise to variations in $g$ values, but likely to a lesser degree.

In addition to the observed variation in coupling strength for different QDs, we also measure slightly different resonance frequencies for each QD due to variations in QD size and shape, which correspond to varying amount of spectral detuning from the plasmon resonance.
Localizing 21 different QDs, we measure their TEPL spectra and determine the corresponding mode energies of the $\ket{\mathrm{UP}}$ and $\ket{\mathrm{LP}}$ branches and the QD energy detuning ($\omega_{SP}$ $-$ $\omega_{QD}$) by fitting to Eq. \ref{Model}.
With the nano-cavity characterized by fixed $\omega_{SP}$ = 1.850 eV and $\Gamma_{SP}$ = 0.160 $\pm$ 0.010 eV and identical for all QDs, and the narrow range of QD linewidth of $\Gamma_{QD}$ = 0.090 $\pm$ 0.010 eV for all QDs, the data can be fit with the QD resonance frequencies $\omega_{QD}$ and the coupling strength $g$ as the only free parameters (see Supplementary Materials V for fitted values). 
The resulting dispersion of the $\ket{\mathrm{UP}}$ and $\ket{\mathrm{LP}}$ branches for each of the 21 QDs are plotted in Fig. \ref{anticrossing}B (circles) as a function of the detuning value, with color scale representing their corresponding $g$ value obtained.
A clear anticrossing is observed as shown in comparison with the expected cavity dispersion curve (red) and average QD energy (blue).

The data scatter in energy is caused by a variation in g from 107 to 152 meV. The corresponding set of plexciton energy dispersions calculated for each QD using Eq. S1 (Supplementary Materials V) are shown in Fig. \ref{anticrossing}B as solid lines. 
This ability to measure individual QDs over a range of detuning values with a single plasmonic nano-cavity demonstrates the distinct advantage of TESC spectroscopy.

\begin{figure}
\center
\includegraphics[width=13cm]{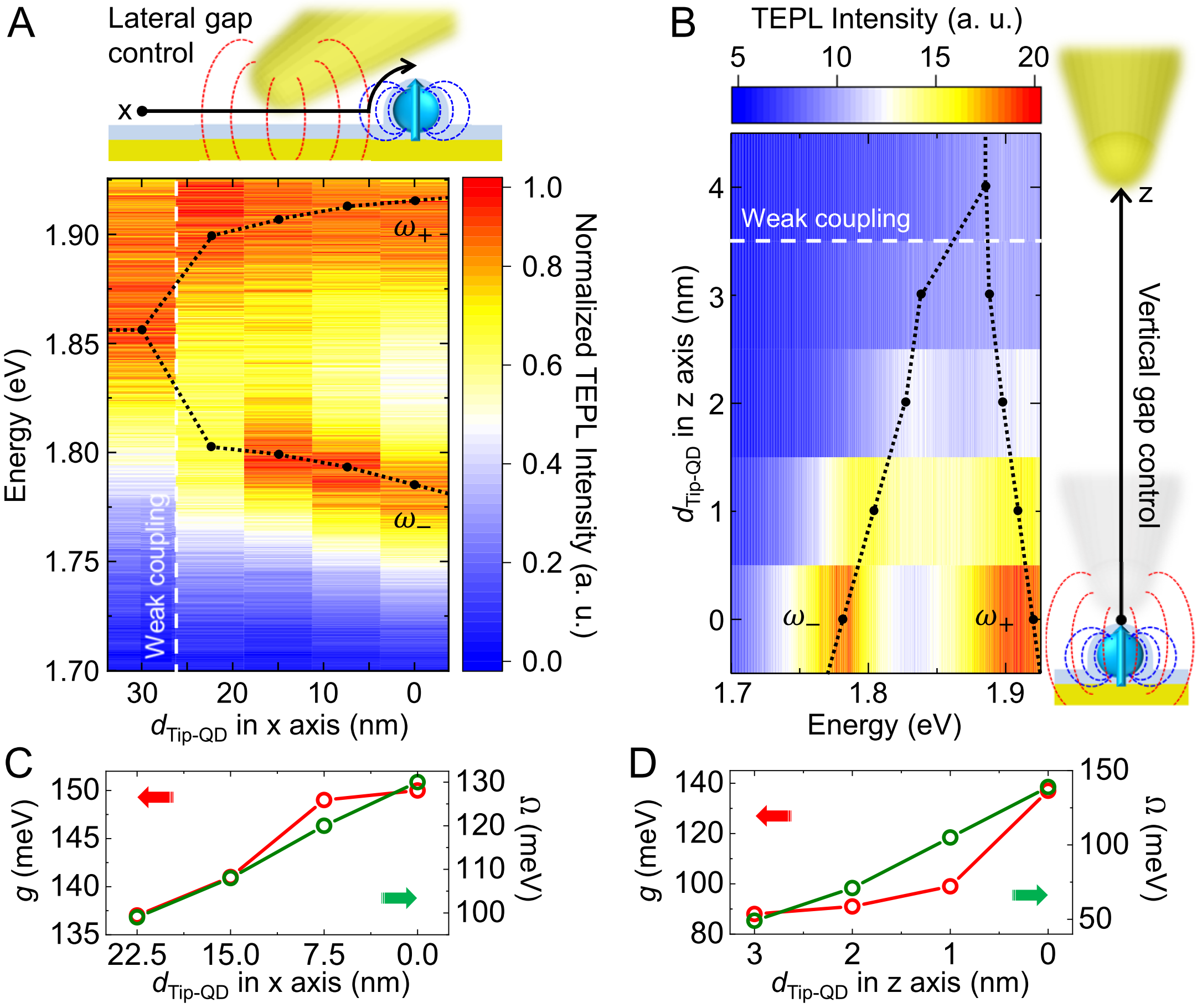}
\caption{
\textbf{Active control of tip-induced single quantum dot strong coupling.} 
TEPL spectra as the lateral (A) and vertical (B) tip-QD distance is varied from 30 nm to 0 nm and from 0 nm to 4 nm, respectively.
(C-D) Coupling strength $g$ and Rabi frequency $\Omega$ derived from model fit of the distance dependent TEPL spectra from (A) and (B). 
}
\label{fig4}
\end{figure}

In addition, TESC provides the ability to tune the coupling strength and control the nano-cavity mode volume in interaction with an individual emitter.
As an example, Fig. 4A shows the transition into the strong coupling regime with a steady increase in Rabi splitting up to $g$ $\sim$ 140 meV as we laterally scan the tip toward a QD over a length scale defined by the tip radius. Conversely, vertical tip-sample distance control with sub-nm precision shown in Fig.~4B reflects the extreme spatial confinement with decreasing gap width \cite{huang2016}. 
The shorter length scale compared to lateral tip-QD separation is due to the rapid decrease in coupling strength $g$.
Fig. 4C-D show the corresponding tip-QD distance dependence of the coupling strength $g$ and Rabi frequency $\Omega$ derived from Fig. 4A-B (for details and corresponding values for $\omega_{SP}$ and $\omega_{QD}$ see Supplementary Materials VI).
Based on these results, the optical mode volume $V/\lambda^3$ of the plasmonic nano-cavity is estimated to be $\leq$ 10$^{-6}$, consistent with the expectation from the $|\textbf{E}_x|$ and $|\textbf{E}_z|$ field distribution calculation ($V$ $\sim 10^3$ nm$^3$), as seen in Fig. 1B-C and can readily be further reduced by using tips of smaller radius \cite{jiang2015, johnson2012}.

It is worth noting that, in contrast to conventional cavities that interact in the far-field, the presence of the emitter strongly modifies the mode volume, polarization distribution, and loss rate of a nano-optical cavity. While this complicates quantification and modeling of these properties, the emitter-cavity backaction provides new control degrees of freedom, and scanning the nano-cavity provides a way to determine its characteristics through TESC imaging.

In the following, we compare our result with recent studies on strong coupling of single emitters using plasmonic nano-cavities. 
To date, most strong coupling studies have relied on multiple ($n$) emitters in the form of J-aggregates \cite{wersall2016} to induce a large collective effect ($g\propto\sqrt{n/V}$). However, multiple emitters are impracticable for applications in quantum gates \cite{reiserer2014} and entanglement \cite{fattal2004} because they exhibit an equidistant energy spectrum that is identical to a classical system. This means that the nonlinear optical properties of multiple emitters are qualitatively different than the single emitter case \cite{fink2008}. However, the observation of peak splitting in scattering spectra from single emitters \cite{chikkaraddy2016, santhosh2016} is not sufficient to ensure strong coupling, as other effects such as Fano-like interference can lead to nearly identical spectral features \cite{leng2018,wersall2016}. Probing strong coupling plexciton emission in the form of, e.g., PL of a single emitter, overcomes these issues and provides for unambiguous evidence of strong coupling for applications in quantum information and photonic quantum devices. However, the recent work using PL to probe strong coupling of a single QD coupled to a plasmonic slot structure \cite{grob2018} led to convoluted spectra with large background and competing interactions. 


In contrast, our approach of TESC employs a combination of several enabling features which allow us to induce, image, and control single emitter plexciton PL at room temperature with the following beneficial attributes. 
First, based on the inverse geometry \cite{yang2014}, instead of a nanogap in the tip itself we induce the nano-cavity between the tip and the sample, which provides for tunable coupling and ultra-small mode volumes (Supplementary Materials VIII).
Second, with the tilted tip and thus spectrally controlled SPP, the field enhancement is further increased by suppressing the overdamped resonance of a conventional surface-normal oriented tip \cite{park2018polarization}.
Third, near-field coupling and polarization transfer to the tip with its few-fs radiative lifetime \cite{kravtsov2014} at the sub-nm gap between the tip and QD outcompetes the otherwise dominant non-radiative damping of a QD near a metallic surface \cite{kongsuwan2017}. 

Single emitters strongly coupled to nano-photonic modes are emerging as a promising enabling technology for nonlinear optics \cite{lukin2000} and quantum information processing \cite{tiecke2014, sun2016} at the single-photon level.
Our demonstration of TESC of single QD plexciton PL opens a new paradigm of near-field microscopy which can be generalized to any optical modality. 
Specifically, the ability to rapidly and reversibly change the coupling strength through nanoscale tip positioning provides a new means of tuning these quantum-optical interfaces, providing a degree of functionality to control quantum dynamics.
TESC can be extended further by using emitters with larger transition dipole moment, tips that provide smaller mode volume, and nano-plasmonic tip engineering to optimize the plasmon resonance. 
It is not limited to plexciton control, and can be applied to a variety of quantum states ranging from exciton-polariton Bose-Einstein condensates \cite{plumhof2013} to infrared vibrational resonances \cite{muller2018}.
Furthermore, it enables novel strong coupling application to chemically diverse species \cite{coles2014, shalabney2015, hutchison2012} to actively and dynamically control photochemical pathways at the single-molecule level.
\\


\bibliography{strong}
\bibliographystyle{customsci}

\section*{Acknowledgments} 
K.-D. Park, M. A. May, J. Wang, and M. B. Raschke acknowledge funding from  the National Science Foundation (NSF Grant CHE 1709822).
H. Leng and M. Pelton acknowledge support from the National Institute of Standards and Technology under Award Number 14D295.
J. A. Kropp and T. Gougousi acknowledge support from the National Science Foundation under grant ECCS-1407677.
{\bf Author contributions:}
M.B.R. and M.P. conceived the experiment.
K.-D.P., M.A.M, and J.W. performed the measurements. 
K.-D.P. performed the FDTD simulations.
H.L. performed the FEM simulations.
H.L., J.A.K., T.G., and M.P. designed and prepared the samples.
K.-D.P., M.A.M., H.L., M.P., and M.B.R. analysed the data, and all authors discussed the results.
K.-D.P., M.A.M., and M.B.R. wrote the manuscript with contributions from all authors. 
M.B.R. supervised the project. 
{\bf Competing interests:}
The authors declare no competing financial interests. 
{\bf Data and materials availability:}
All data are presented in the main text and supplementary materials.

\section*{Supplementary Materials}
Materials and Methods\\
Supplementary Text\\
Figs. S1 to S4\\


\end{document}